\documentclass[journal]{IEEEtran}
\usepackage[utf8]{inputenc}
\usepackage{caption}
\usepackage{multicol}
\usepackage{multirow}
\usepackage{amsmath,amssymb,amsfonts}
\usepackage{algorithmic}
\usepackage{graphicx}
\usepackage{balance}
\usepackage{textcomp}
\usepackage[dvipsnames]{xcolor}
\usepackage[utf8]{inputenc}
\usepackage[perpage,symbol]{footmisc}
\usepackage[english]{babel}
\usepackage{amsmath, bm}
\usepackage{pbox, booktabs, rotating}
\usepackage{subfigure}
\usepackage{array, makecell, lipsum}
\usepackage{enumerate}
\usepackage{threeparttable}
\usepackage{multirow,hyphenat,balance}
\usepackage{placeins}
\usepackage[compress]{cite}
\usepackage{float,color}
\usepackage{ragged2e}
\usepackage{enumerate}
\usepackage{etoolbox}
\usepackage{paralist}
\usepackage{cases}
\usepackage{url}
\usepackage{framed}
\usepackage{flushend}
\usepackage{amsmath}
\usepackage{amsthm}
\usepackage{amsthm}
\usepackage[english]{babel}
\usepackage{amssymb}
\usepackage[caption=false,font=normalsize,labelfont=sf,textfont=sf]{subfig}
\usepackage[linesnumbered,ruled]{algorithm2e}
\let\oldnl\nl
\newcommand{\nonl}{\renewcommand{\nl}{\let\nl\oldnl}}
\usepackage{pifont}
\usepackage{commath}
\usepackage{flushend} 

\usepackage{mathtools}

\newcolumntype{P}[1]{>{\centering\arraybackslash}p{#1}}
\newcolumntype{M}[1]{>{\centering\arraybackslash}m{#1}}

\newcommand*{\el}{et al.\@\xspace}

\DeclareCaptionLabelFormat{nospace}{#1#2}
\def\paragraph#1{\textbf{\color{orange}#1}}

\title{Advancing Security with Digital Twins: A Comprehensive Survey}
\author{Blessing Airehenbuwa, \IEEEmembership{Student Member IEEE}, Touseef Hasan, \IEEEmembership{Student Member IEEE}, Souvika Sarkar, \IEEEmembership{Member IEEE} and Ujjwal Guin, \IEEEmembership{Senior Member IEEE}

\thanks{Blessing Airehenbuwa and Ujjwal Guin are with the Department of Electrical and Computer Engineering, Auburn University, AL, USA (e-mail: \{b.airehenbuwa and ujjwal.guin\}@auburn.edu).}
\thanks{Touseef Hasan and Souvika Sarkar are with the School of Computing, Wichita State University, KS, USA (e-mail: \{txhasan4@shockers.wichita.edu and souvika.sarkar@wichita.edu\}).}
}

\begin{document}

\IEEEoverridecommandlockouts

\maketitle

\begin{abstract}
The proliferation of electronic devices has greatly transformed every aspect of human life, such as communication, healthcare, transportation, and energy. Unfortunately, the global electronics supply chain is vulnerable to various attacks, including piracy of intellectual properties, tampering, counterfeiting, information leakage, side-channel, and fault injection attacks, due to the complex nature of electronic products and vulnerabilities present in them. Although numerous solutions have been proposed to address these threats, significant gaps remain, particularly in providing scalable and comprehensive protection against emerging attacks. Digital twin, a dynamic virtual replica of a physical system, has emerged as a promising solution to address these issues by providing backward traceability, end-to-end visibility, and continuous verification of component integrity and behavior. In this paper, we present a comprehensive survey of the application of digital twins based on their functional role and application domains. We comprehensively present the latest digital twin–based security implementations, including their role in cyber-physical systems, Internet of Things, cryptographic systems, detection of counterfeit electronics, intrusion detection, fault injection, and side-channel leakage. This work comprehensively considers these critical security use cases within a single study to offer researchers and practitioners a unified reference for securing hardware with digital twins. Additionally, we discuss the integration of large language models into digital twins to enhance the security of systems by leveraging their advanced reasoning capabilities and highlight the challenges and limitations of applying digital twins to solve hardware security problems, with possible solutions. Further, we discuss current research trends and future directions in advancing hardware security through digital twins.
\end{abstract}

\vspace{5px}


\maketitle

\section{Introduction}\label{sec:intro}
The modern electronics supply chain is a highly intricate and globalized ecosystem that spans across multiple countries and involves various independent entities responsible for different stages of the product lifecycle, including design, fabrication, packaging, assembly, and distribution~\cite{guin2013anti, tehranipoor2015counterfeit}. This complexity is compounded by the presence of both trusted and untrusted actors, increasing the risk of vulnerabilities being introduced at any point in the supply chain~\cite{guin2014counterfeitProc}. The cross-border nature of operations makes oversight and accountability particularly challenging, as different regions may follow varying standards of security, regulation, and quality assurance. Furthermore, the supply chain consists of multiple hierarchical layers, often including subcontractors and third-party vendors, which makes it difficult for original equipment manufacturers to maintain complete visibility into every component's origin and traceability information. A significant limitation is the lack of real-world implementation of a tracking system that provides the ability to trace components back to their source and verify their integrity~\cite{valapu2024reward,zhong2023blockchain, cui2019blockchainProvenance}. This lack of transparency and visibility not only complicates risk management but also creates opportunities for adversaries to insert counterfeit parts, malicious circuitry, or compromise intellectual properties. As the demand for faster, cheaper, and more powerful electronics grows, addressing these challenges becomes critical to ensure the security, reliability, and trustworthiness of electronic systems deployed across commercial, industrial, and national security applications.

Although solutions exist to address these security challenges -- such as counterfeit ICs~\cite{guin2014counterfeitProc, tehranipoor2015counterfeit, guin2014counterfeitJETTA}, information leakage~\cite{hu2021hardware, rajendran2016formal, hovanes2022beware}, hardware Trojans~\cite{karri2010trustworthy, jain2021survey, tehranipoor2014integrated}, and side-channel attacks~\cite{kocher1999differential, standaert2010introduction, schneider2015leakage}. However, research gaps still exist as these approaches often focus on a specific threat, while the attack surface continues to expand due to the constant development of new threats by adversaries~\cite{jin2015introduction,shaikh2022digital}. Therefore, there is a need for comprehensive and scalable solutions for adapting to emerging threats. A promising solution to address these security concerns can be the adoption of digital twin (DT) technology, which creates a virtual replica of the physical object, providing real-time monitoring, forecasting, simulation, and analysis of each step in the manufacturing and delivery processes.

With these abilities, DTs are able to capture and analyze data throughout the lifecycle of electronics to detect anomalies indicative of counterfeit components. They support robust security analysis and real-time threat detection by continuously evaluating system behavior against expected virtual models. In addition, they facilitate the simulation and emulation of physical systems within a virtual environment, allowing for the identification and mitigation of potential leakage points in designs early in the development phase to improve the security of the system. This paper, therefore, demonstrates the current status of digital twin-enabled hardware security by reviewing relevant research from recent years. To contextualize this survey within existing literature, a comparative overview of related works is provided in Table~\ref{tab:comparisons}. While prior surveys have focused on specific domains such as CPS~\cite{eckhart2019digital, pokhrel2020digital}, or IoT applications~\cite{el2024systematic}. In contrast, this survey offers a broader and unified perspective by covering CPS, IoT, supply chain, and cryptographic systems in a single study, while also examining the integration of LLMs with DT for intelligent security analysis.

\begin{table}[h]
\centering 
\caption{Positioning this survey among related works.} 
\label{tab:comparisons} 
\begin{tabular}{|P{.6cm}|P{.7cm}|P{.4cm}|P{.5cm}|P{.9cm}|P{.9cm}|P{.6cm}|}
\hline
\multirow{3}{*}{\textbf{Year}} & \multirow{3}{*}{\textbf{Ref.}} & \multicolumn{5}{c|}{\textbf{Domains Covered}} \\
\cline{3-7}
& & \textbf{IoT} & \textbf{CPS} & \textbf{Supply Chain} & \textbf{Crypto Systems} & \textbf{LLM} \\
\hline
2019 &~\cite{eckhart2019digital} &  & $\checkmark$ &  &  &  \\
\hline
2020 &~\cite{pokhrel2020digital} &  & $\checkmark$ &  &  &  \\
\hline
2024 &~\cite{el2024systematic} & $\checkmark$ & $\checkmark$ &  &  &  \\
\hline
-- & This Study & $\checkmark$ & $\checkmark$ & $\checkmark$ & $\checkmark$ & $\checkmark$ \\
\hline
\end{tabular}  
\end{table}


The main contribution of our work is to provide the history and fundamental concept of digital twins, a comprehensive review of DT applications in hardware security, and future research directions. Key contributions of this work include:

\noindent \textbullet~ \textit{Application of Digital Twins:} We present detailed applications of DT across multiple domains, including manufacturing, smart cities, healthcare, and energy, highlighting how DTs enable real-time monitoring, predictive maintenance, performance optimization, and enhanced decision-making. Although few surveys exist in the literature~\cite{rasheed2020digital,pokhrel2020digital,rathore2021role}, this survey focuses primarily on the security aspects.

\noindent \textbullet~ \textit{Digital Twin for Security Enhancement:} We present a comprehensive survey of the latest DT–based security implementations and applications spanning both counterfeit detection and information‑leakage prevention in electronics. This work comprehensively considers critical security use cases within a single study to offer researchers and practitioners a unified reference for safeguarding electronic hardware with digital twins. 
We also discuss the use of DTs to enhance security in IoT and cyber-physical systems.
    
\noindent \textbullet~ \textit{Large Language Models in Digital Twins:} The paper examines how the integration of large language models (LLMs) into DT frameworks strengthens hardware security via natural-language understanding and advanced reasoning. Such LLM-enhanced DTs have the potential to enable conversational and interactive interfaces, generate detailed scene descriptions, and provide evidence-based decision-making. These features ensure secure and resilient operations across diverse applications, including hardware security.
    
\noindent \textbullet~ \textit{Future Directions:}  We outlined the current challenges and limitations associated with applying DTs in hardware security, as identified in prior works~\cite{shaikh2022digital, huang2020blockchain, putz2021ethertwin, son2022design}, along with proposed solutions to address them. Recent studies employing LLMs for security verification~\cite{meng2023unlocking, fu2023llm4sechw} have demonstrated significant potential. Since DTs and LLMs individually enhance security, we envision that their convergence will play a pivotal role in shaping the future of security solutions. Although the integration of DTs with LLMs is currently in the research phase within non-security domains, it has yet to be explored for security.

The remainder of this paper is organized as follows. In Section~\ref{sec:DT}, we present an overview of digital twins. We investigate the applications of digital twins in different domains in Section~\ref{sec:application}. Section \ref{sec:DT-in-HS} discusses digital twins for ensuring the security of electronic products. In Section \ref{sec:llms}, we discuss the use of large language models in DT frameworks. We summarize the challenges and limitations of digital twins in Section \ref{sec:DT-challenges}. 
Section \ref{sec:trens} summarizes the research trend. Finally, we conclude the paper in Section~\ref{sec:conclusion}.

\section{Digital Twin} \label{sec:DT}
Digital Twin has received different definitions by different authors over the years based on their perspectives of the concept~\cite{tao2018digital} and applications. According to Glaessgen \el in 2012, a digital twin is an integrated multiphysics, multiscale, probabilistic simulation of a system that uses the best available physical models, sensor updates, history, etc., to mirror the life of its corresponding twin~\cite{glaessgen2012digital}. Years later, in 2017, Grieves and Vickers defined it as a set of virtual information constructs that fully describe a potential or actual physical manufactured product from the micro-atomic level to the macro-geometrical level. Therefore, a digital twin is a dynamic or living virtual replica of a physical object, system, or process. An important thing to note is that the digital twin evolves continuously by integrating real-time data from sensors, IoT devices, and other data sources, to ensure the virtual twin stays in sync with its physical counterpart throughout its lifecycle. The real-time synchronization enables the virtual twin to reflect the current state of its physical counterpart at any given time, thereby providing remote access to the status and conditions of the physical system from anywhere around the globe~\cite{grieves2011virtually}. In addition to the aforementioned benefit, it also provides intelligent feedback (e.g., forecasting, optimization of parameters, root cause analysis, real-time control) to the physical world through a combination of simulation, emulation, data analytics, and AI modeling~\cite{shaikh2022digital}. 

Although DT was originally introduced for PLM~\cite{grieves2006product}. However, we have in recent times seen its application in various fields such as in agriculture~\cite{madjid2024design,pylianidis2021introducing,angin2020agrilora}, healthcare~\cite{liu2019novel}, manufacturing~\cite{xu2019digital,booyse2020deep}, etc. In healthcare, based on real-time health data (e.g., vital signs, medical history, and test results) available, healthcare professionals can use digital twins to simulate how a patient might respond to different treatments, medications, or surgical procedures and predict potential complications, provide personalized treatment plans, and improve diagnostic accuracy without putting the patient at risk due to unnecessary tests and procedures \cite{ankitha2024healthcare}. In a manufacturing setting, a digital twin provides predictive maintenance capabilities by detecting anomalies in machines from operational data early before they lead to failures, thereby reducing downtime and repair costs. 

\subsection{History of Digital Twins}
In 2002, Michael Grieves introduced the concept of digital twin during a University of Michigan presentation to industry for the formation of a Product Lifecycle Management center in what was referred to as ``Conceptual Ideal for PLM''. Although the idea of twinning things can be traced back to NASA's Apollo programs in the 1960s, where identical space vehicles were built to mirror the conditions of the space vehicle during space missions~\cite{ferguson2020apollo, allen2021digital}. During flight missions, another physical vehicle (a twin) remained on Earth and was used to mirror the flight conditions as precisely as possible, based on available flight data. The physical twin on Earth was used to simulate ways to assist crew members onboard in critical situations. This groundbreaking initiative laid the foundation for what would later become a widely recognized framework, leveraging the advanced technological and scientific methodologies developed to ensure mission success and innovation in space exploration. However, building such physical twins has become increasingly expensive as the cost of physical materials (atoms) has continued to increase. Hence, the need to create and operate the twin in a virtual or digital space was proposed by Grieves.

Since its conceptual introduction in 2002, it has been referred to by different names over the years. In 2005, Grieves referred to it as the Mirrored Spaces Model, which he later referred to as the Information Mirroring Model in 2006~\cite{grieves2006product}. However, this concept was significantly expanded on in his 2011 work~\cite{grieves2011virtually}, where the term ``Digital Twin'' was used for the first time, a name coined by John Vickers, NASA. While the name has changed over the years, the concept and model have remained the same.
DT has gained applications and recognition in various fields. In 2010, NASA included the technology in their technology roadmap~\cite{piascik2010technology} and explored its potential application in spacecraft~\cite{glaessgen2012digital}. Similarly, the U.S. Air Force leveraged the concept for jet fighters, as highlighted by Tuegel~\el~\cite{tuegel2011reengineering}. This marked a pivotal moment in the evolution of digital twin technology, showcasing its growing importance in advanced applications.

\subsection{Concept of Digital Twins}
\begin{figure}[!ht] \vspace{-10px}
    \centerline{\includegraphics[width=1\linewidth]{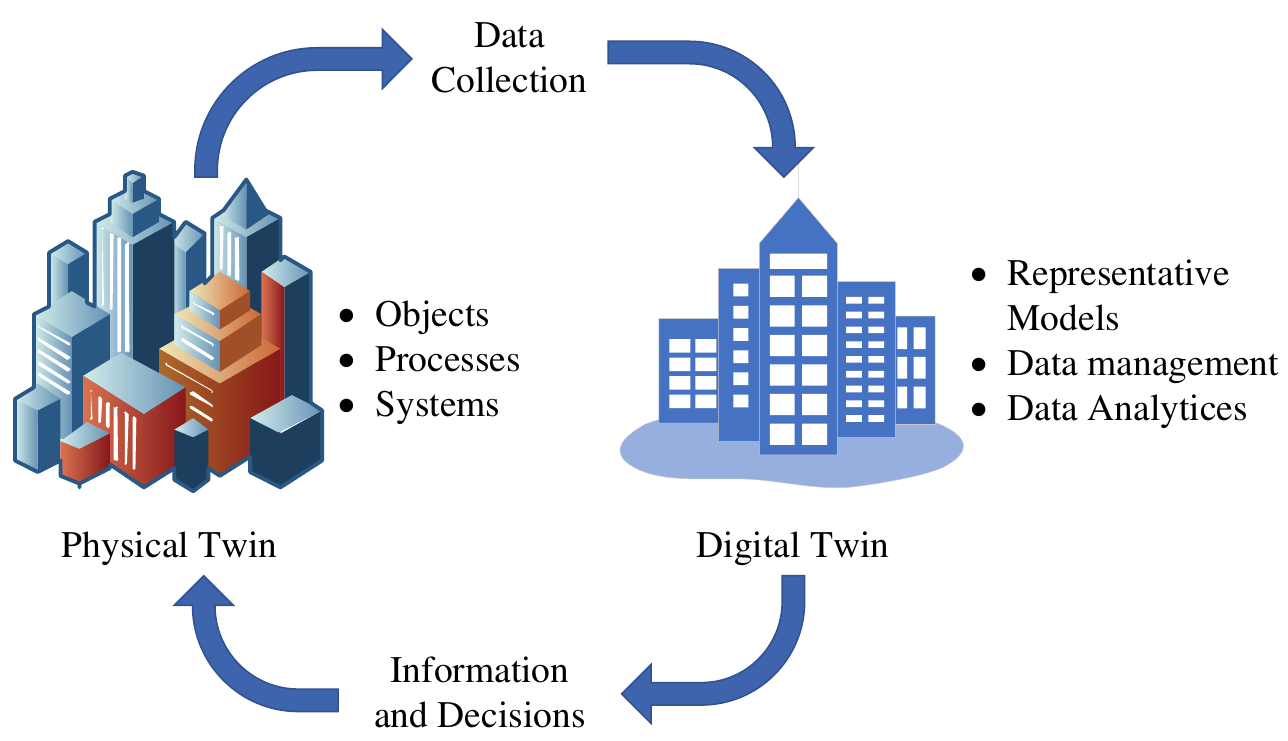}}
    \caption{\centering The conceptual diagram of a digital twin~\cite{gao2023digitaltwins}.} \vspace{-10px}
    \label{fig:DT}
\end{figure}

The basic idea of the digital twin is to represent a physical product, system, or process in the virtual domain to monitor, simulate, and optimize its performance through real-time data integration. By doing so, we can obtain the same information from the virtual twin as could have been obtained if we had access to the physical object. In addition, we can perform what-if analysis on the virtual twin without putting the physical counterpart at risk. In principle, a digital twin consists of three (3) components, namely, the physical twin in the real space, the virtual twin, and the communication links between them~\cite{grieves2017digital,grieves2023digital}. Figure~\ref{fig:DT} shows a conceptual diagram of a digital twin~\cite{gao2023digitaltwins}.

The physical twin is the real-world product, process, system, or asset being digitally represented. It serves as the foundation or source for the digital twin's data and functionality. This component of the digital twin resides in real space and interacts with its environment, performs its designated tasks, and undergoes wear and tear. All of these provide the essential data needed for the digital twin to simulate, analyze, and predict future behaviors and outcomes. On the other hand, there is the virtual or digital twin, which is a representation of the physical counterpart in the virtual space or environment, and is capable of emulating the behavior of its physical twin in great detail. The virtual twin can be modeled as a physics-based model from first principles. However, obtaining a physics-based model may be infeasible for some complex systems. In such situations where the physics-based models are not easily available, data-driven models are used. This approach is based on the assumption that data is a manifestation of both known and unknown physics~\cite{rasheed2020digital}. To build this type of model, one is required to perform data collection, data pre-processing, and data analysis with Artificial Intelligence (AI). An example of this type of model is found in~\cite{booyse2020deep,xu2019digital}. Sometimes, a combination of physics-based and data-driven models is required to model the system. This type of digital twin model is referred to as the hybrid model~\cite{segovia2022design,willard2022integrating}. A graphical approach to modeling a system was discussed in~\cite{kapteyn2021probabilistic}. In short, the virtual twin is the component of the digital twin that enables detailed simulations, optimization, diagnosis, and predictions to support decision-making. Lastly, the connection link is the component of digital twins that allows data to flow from the physical twin to the virtual twin in the form of sensor data and historical data. Depending on the system being modeled, information and decisions will flow from the virtual space to the real space.

\begin{figure*}[!ht] 
    \centering
    \includegraphics[width=1\textwidth]{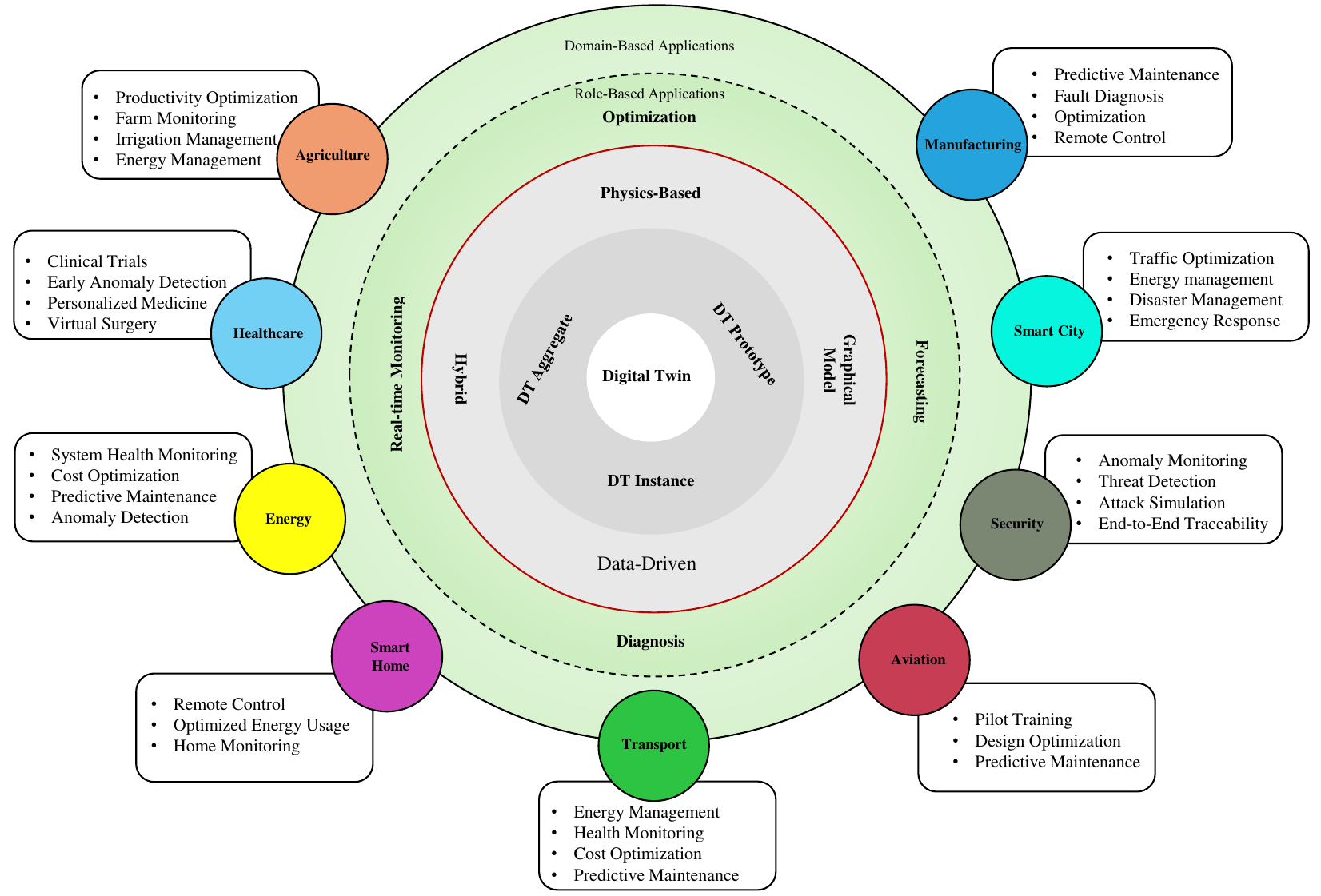} 
    \caption{\centering Digital Twin Taxonomy and Applications.} 
    \label{fig:DT-Tax}
\end{figure*}

To better understand digital twins, we present a comprehensive and visualized taxonomy of digital twins in Figure~\ref{fig:DT-Tax} based on their types and virtual twin modeling methods. We also present the applications of digital twins based on their role and domain of application in the outer layers. 

Digital twins are classified into three types~\cite{grieves2017digital} -- digital twin prototype, digital twin instance, and digital twin aggregate, shown in the second ring of Figure~\ref{fig:DT-Tax} and described in Section~\ref{sec:DT-types}. The third ring describes the virtual twin modeling methods, which are grouped into physics-based, data-driven, hybrid, and graphical methods. In the fourth and fifth rings, we present the applications of digital twins based on their role and the domain applied, respectively. These applications are described in Section~\ref{sec:application}.

\subsection{Types of Digital Twins}\label{sec:DT-types}
Digital twins are categorized into three main types, each serving a distinct role throughout a product’s lifecycle. From the initial design phase to real-time monitoring and large-scale analysis. According to~\cite{grieves2017digital}, the three types of digital twins are as follows:

\vspace{2px}
\noindent \textbullet~ \textit{Digital twin prototype (DTP)} represents a product in its pre-manufacturing stage. It includes all the necessary information to design, produce, and describe a physical counterpart, such as requirements, fully annotated 3D models, bill of materials (BoM), and bills of processes, services, and disposal.

\vspace{2px}
\noindent \textbullet~ \textit{Digital twin instance (DTI)} represents the digital counterpart of an individual manufactured product. It is created when the physical product rolls off the production line and is linked to its corresponding physical twin throughout its lifecycle. This connection allows for real-time synchronization and updates, enabling effective monitoring and management of the product’s performance, maintenance, and overall lifecycle.

\vspace{2px}
\noindent \textbullet~ \textit{Digital twin aggregate (DTA)} aggregates multiple DTIs, providing a composite view of all instances of a product or system. This aggregation facilitates large-scale analysis, such as identifying trends, optimizing fleet performance, and improving operational efficiency across multiple units.

\section{Application of Digital Twin} \label{sec:application}
In this section, we categorize the applications of digital twins and briefly discuss them along two dimensions: first, based on the functional role of the digital twin, and second, based on the domain in which it is applied.

\subsection{Role-Based Application of DT}
Depending on the application, a DT performs any or a combination of these functions:

\vspace{2px}
\noindent \textbullet~\textit{Real-time Monitoring:} This is the simplest function of a digital twin. Here, the digital twin simply reflects the state of the physical system at any point in time based on real-time data collected from the physical system. With this, no matter where the physical system is in the world, we can get information about its state and performance. For instance, we can monitor the health of industrial machines, bridges, vehicles, electronics, etc. A practical example is found in~\cite{mihai2022digital}, where a digital twin was used to monitor the Structural Health of bridges in Vietnam. The design incorporated sensors to collect operational data, and a fog layer for local pre-processing before transmitting the data to the cloud for further analysis and visualization. Similarly, Li \el introduced a cloud-based Battery Management System (BMS) to build a digital twin of battery systems, ensuring accurate status monitoring~\cite{li2020digital}. Another application of digital twin for real-time monitoring can be found in~\cite{li2024data}, where Li \el proposed a technique to monitor the health and performance of a boost converter in power electronics. In summary, as DT continues to evolve, its real-time monitoring applications will play an increasingly important role in optimizing performance and decision-making across various domains.
    
\vspace{2px}
\noindent \textbullet~\textit{Optimization:}\label{Dt-subfunc-opt} With digital twin, we have access to real-world data from the physical system, the virtual twin, and AI. The integration of these components for real-time simulations allows engineers to explore multiple design configurations and optimize critical parameters to improve the performance of the system.
Let us consider a scenario such as the design of an aircraft. With the digital twin of the aircraft, we can simulate airflow, thermal stresses, and component interactions to refine the designs, leading to an optimal design that reduces drag and improves engine efficiency. Similarly, in manufacturing, digital twins have the potential to simulate material flow, machine-to-machine interactions, and energy consumption to optimize production layouts and waste. The application of digital twin for optimization can be found in ~\cite{tao2017digital,tao2018digital},  where the Digital Twin Shop Floor (DTS) was proposed to optimize manufacturing through an iterative process that integrates real-time data from the Physical Shop Floor (PS), Virtual Shop Floor (VS), and Shop Floor Service System (SSS). The digital twin detects potential resource allocation conflicts and provides optimization strategies to enhance efficiency before actual execution. During manufacturing, PS transmits the real-time state data to VS, which updates itself to reflect the physical changes, ensuring alignment with predefined plans. This continuous feedback loop minimizes disruptions, enhances precision, and maximizes overall operational efficiency, making DTS a powerful tool for intelligent and adaptive manufacturing optimization.
    
We can also find the optimization capability of digital twins as proposed by Bellalouna \cite{bellalouna2021case}. The author used real-time sensor data and advanced processing technologies to precisely determine material stress and load conditions to facilitate efficient product design as opposed to traditional mechanical structure analysis, which relied on estimated data and theoretical calculations, often leading to over-dimensioning and material waste due to high safety factors. In summary, design teams can use digital twins to identify and mitigate inefficiencies, minimize production costs, and accelerate time-to-market while ensuring the highest quality and performance.

\vspace{2px}
\noindent \textbullet~\textit{Forecasting:} Digital twin serves as a predictive tool to forecast the future behavior and performance of the physical twin. The digital twin achieves this function by using its AI component to analyze the observed measurements coming from sensors attached to the physical twin. With this, we could predict when a system or its subsystem will fail. Thereby eliminates the unexpected failure of the system. It is important to know that this function of digital twins is the reason for the widespread adoption of digital twins in manufacturing in recent years, to reduce operational downtime. For example, in a recent study, Aivaliotis~\el~demonstrated the use of digital twins to predict the health condition of a six-axis robotic structure used for welding in order to reduce downtime~\cite{aivaliotis2019use}. Similarly, the authors used digital twins to predict the health status and maintenance cycles of machines~\cite{centomo2020design}. Similarly, \cite{li2024data, li2020digital, xiong2021digital} employed digital twins to predict the remaining useful life of systems. 

\vspace{2px}
\noindent \textbullet~\textit{Diagnosis:} Digital twin technology is a transformative leap in predictive maintenance and fault detection, enabling engineers to detect faults~\cite{jain2019digital} and also trace the origin of such faults from the observed historical data of a physical system \cite{shaikh2022digital}. By employing advanced analytics, machine learning (ML), and real-time data integration, the diagnosis function of digital twins empowers industries to transition from reactive to proactive maintenance strategies. Ultimately, improves systems' reliability, sustainability, and performance.

\subsection{Domain-Based Application of Digital Twins}\label{sec:domain-based-dt}
Here, we provide a comprehensive summary of the literature on the application of digital twins across various domains such as agriculture, healthcare, smart homes, manufacturing, transportation, energy, and aviation.

\vspace{2px}
\noindent\textbullet~\textit{Agriculture:} In agriculture, digital twins enable real-time monitoring of farms and greenhouses by tracking soil moisture, temperature, crop health, and weather~\cite{madjid2024design,lakhiar2018monitoring,vicente2014dynamical}. It allows resource optimization as demonstrated in Manocha~\el~work, where DT was used to optimize water usage during irrigation, crop stress prediction~\cite{moghadam2020digital}, and climate control optimization for indoor agriculture~\cite{chaux2021digital}.

\begin{figure*}[!ht]
    \centerline{\includegraphics[width=0.93\linewidth]{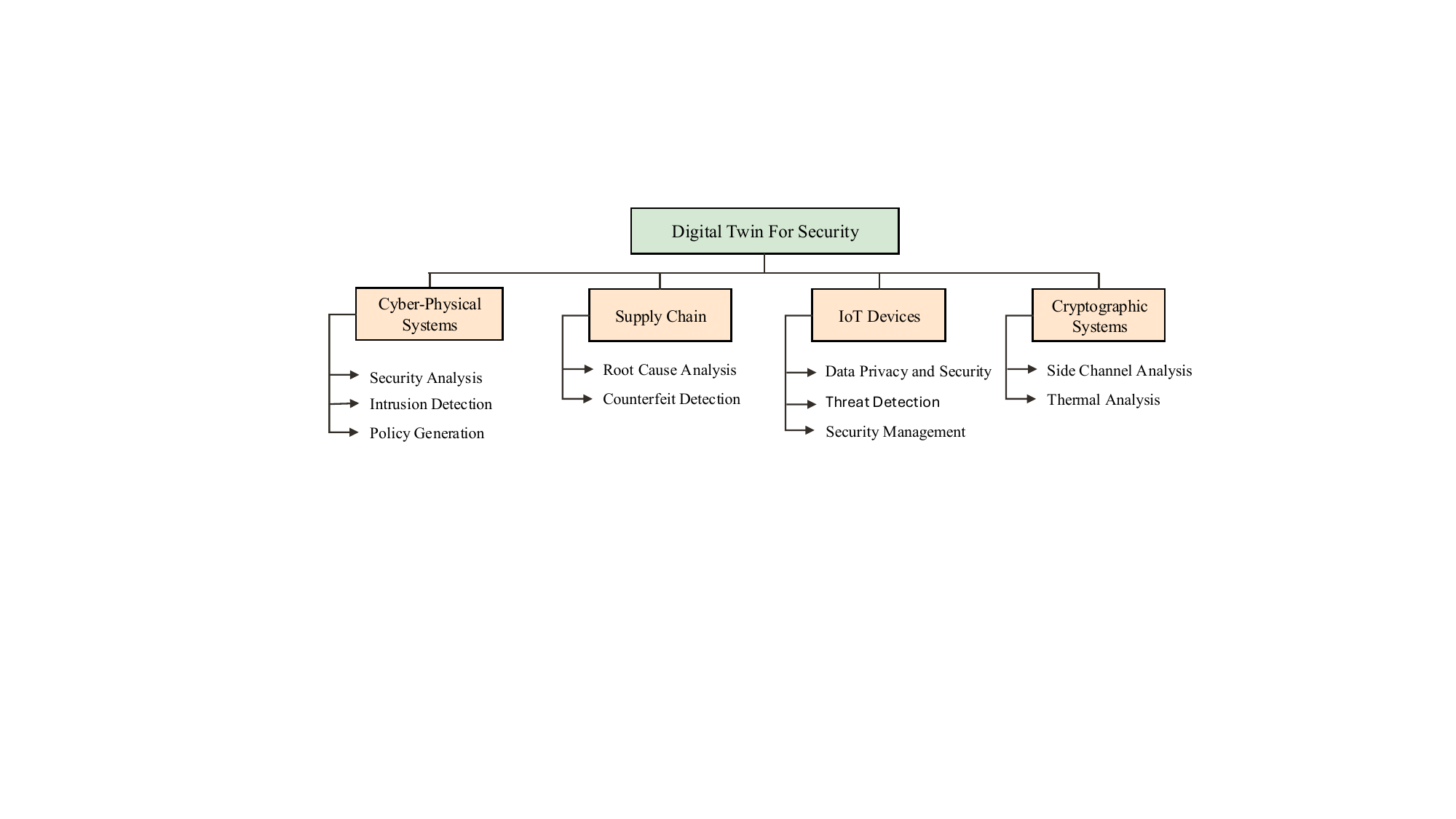}} 
    \caption{\centering A taxonomy of applications of digital twins for enabling security. 
    }
    \label{fig:HWSec_Tax}
\end{figure*}

\vspace{2px}
\noindent  \textbullet~\textit{Aviation:}
Digital twins advance aviation through predictive maintenance, health monitoring, and anomaly detection to improve safety and efficiency. In~\cite{kapteyn2021probabilistic}, Kapteyn~\el~demonstrated the effectiveness of digital twins in mission planning and UAV fleet optimization. It also finds usefulness in aircraft damage prediction by leveraging flight data and Bayesian learning~\cite{tuegel2011reengineering}. Similarly, Wang~\el~proposed an approach for diagnostics and prognostics of damaged aircraft by combining fatigue mechanics and filtering theories~\cite{wang2015use}. Another application of digital is found in the work of Zakrajsek~\el, which demonstrated the use of digital twins to predict wear in aircraft tires under different touchdown conditions.

\vspace{2px}
\noindent \textbullet~\textit{Energy:} 
DTs find several applications in the energy domain, such as real-time monitoring, optimization, and predictive maintenance. A typical example in this domain is General Electric's digital wind farm, which uses DTs for real-time turbine monitoring and predictive maintenance~\cite{lund2018digital}. In~\cite{li2024data,wunderlich2021digital}, the authors utilized DTs to detect degradation, fault, prognosis, and health management~\cite{wunderlich2021digital}. Saad~\el~used IoT-based DTs to secure microgrids against cyberattacks, improving system resilience~\cite{saad2020implementation}. Broader DT applications in power systems are detailed in~\cite{jain2019digital,saad2020iot,pan2020digital,song2020parameter,saad2018data}.

\vspace{2px}
\noindent  \textbullet~\textit{Healthcare:}
In healthcare, digital twins support personalized treatment by replicating patient physiology~\cite{erol2020digital}. Beyond personalized treatment, it also enables virtual medical training~\cite{haleem2023exploring}, health monitoring, and diagnosis, such as CloudDTH~\cite{liu2019novel}. Another application of DT in healthcare is found in~\cite{chen2023digital}, detecting atrial fibrillation and falls using ECG and WiFi signal.

\vspace{2px}
\noindent  \textbullet~\textit{Manufacturing:} 
Digital twin is a key enabler of Industry 4.0, supporting real-time monitoring, fault diagnosis, optimization, and enhanced security. An example of this application is discussed in~\mbox{\cite{tao2017digital, tao2018digital}}, where the authors proposed a Digital Twin Shopfloor (DTS) which integrates physical, virtual, and service systems for adaptive control at the shop floor. Xu~\mbox{\el} introduced a DT-based fault diagnosis framework using synthetic and real data for early fault detection~\mbox{\cite{xu2019digital}}. Booyse~\mbox{\el} developed Deep DTs to detect degradation from healthy data~\mbox{\cite{booyse2020deep}}. To enhance the security of systems, Salim~\mbox{\el} proposed a blockchain-enabled digital twin that uses smart contracts to detect botnets while preserving data privacy~\mbox{\cite{salim2022blockchain}}. Becue~\mbox{\el} applied DTs to evaluate attack impacts and strengthen system resilience~\mbox{\cite{becue2018cyberfactory}}.

\vspace{2px}
\noindent \textbullet~\textit{Smart Home:}
DTs enhance smart homes by leveraging IoT sensors, ML, and actuators to improve comfort, security, and energy efficiency. Liu~\mbox{\el} proposed a DT framework for indoor safety using IoT sensor data~\mbox{\cite{liu2020framework}}, while Khajavi~\mbox{\el} applied DTs for safety and fire detection~\mbox{\cite{khajavi2023digital}}. Gopinath~\mbox{\el} redesigned smart homes with DTs, enabling remote control of appliances and real-time energy monitoring~\mbox{\cite{gopinath2019re}}. DTs also support virtual testing and reduce downtime, boosting system reliability and energy efficiency~\mbox{\cite{fujii2075digital, gopinath2019re}}.

\vspace{2px}
\noindent\textbullet~\textit{Transportation: } 
Kumar~\mbox{\el} introduced an AI-driven traffic management system using digital twins, blockchain, LSTM models, and edge/fog computing to predict driver behavior and optimize routes~\mbox{\cite{kumar2018novel}}. In another work, Ye~\mbox{\el} used DT-enhanced Q-learning to simulate driving conditions and battery degradation, improving energy efficiency~\mbox{\cite{ye2022reinforcement}}. Venkatesan~\mbox{\el} proposed an intelligent DT for electric motor health monitoring using ANN and fuzzy logic, combining onboard diagnostics with cloud-based prognosis~\mbox{\cite{venkatesan2019health}}.
For traffic modeling, Ku{\v{s}}i{'c}~\mbox{\el} developed the Geneva Motorway DT (DT-GM), which integrates real-time data with SUMO simulations and is updated every minute to reflect the dynamic highway conditions~\mbox{\cite{kuvsic2023digital}}. From a security perspective, Arya~\mbox{\el} developed a DT framework with machine learning to detect malicious nodes and mitigate DDoS attacks in vehicular networks~\mbox{\cite{arya2022detection}}. Tesla also currently utilizes digital twins to enable predictive maintenance, over-the-air updates, and real-time analytics~\mbox{\cite{mike2024tesla}}.

\section{Digital Twin for Security} \label{sec:DT-in-HS}
The DT framework presents itself as the ideal tool to enhance security in electronic devices due to its inherent abilities in detecting anomalies (e.g., vulnerabilities/flaws in design) from data, and the formulation of robust security strategies through playing what-if scenarios.
The potential of DT in this domain has been widely acknowledged by security researchers, hence the need for this study~\cite{eckhart2018towards, Hornik2024codt, benelhaouare2024enhancing, shaikh2022digital, bohm2021augmented, sarker2024explainable, hammar2023digital}. 
In this section, we present recent studies on the exploration of DT for securing electronic systems. 
Figure~\ref{fig:HWSec_Tax} shows how DTs have been employed to address security concerns, including cyber-physical systems, counterfeit electronics, IoT devices, and cryptographic systems, and is described in the following.

\subsection{Cyber-Physical Systems}

DTs are employed as a cutting-edge tool to address security challenges due to their predictive and diagnostic strengths. DTs provide security experts with valuable benefits, providing the ability to detect, predict, simulate, and test cyber-attacks without interacting with the physical system~\cite{dietz2020unleashing,eckhart2018towards}. Several approaches have recently been proposed in this domain, utilizing DTs to detect and prevent threats and attacks. The goal is for cyber threats to be predicted and mitigated before they occur, ensuring that the physical system remains secure and updated on potential risks. In this work, we categorize the application of DT in cyber-physical systems into three key areas: security analysis, intrusion detection, and policy generation, each of which is discussed below.

\vspace{5px}
\subsubsection{Security Analysis}
Digital twins have emerged as a powerful tool for conducting security analysis in cyber-physical systems, particularly where direct testing on live systems is impractical or risky. Security testing in CPS is critical due to their exposure to real-world cyber threats; however, performing such analysis on live systems can compromise their operation and safety. DTs overcome this problem by offering a virtual replica of the physical system, allowing researchers to simulate attacks, analyze vulnerabilities, and develop mitigation strategies in a safe and cost-effective environment. This approach was employed by Shitole~\el where a low-cost, real-time DT of interconnected residential energy storage systems was introduced to enable accurate security analysis and intrusion detection without modifying or risking the physical system~\cite{shitole2021real}. 
In another work, Dietz~\el developed a process-based security framework to incorporate digital twin security simulations in the Security Operations Center (SOC). The authors demonstrated a man-in-the-middle attack and analyzed the effect using the virtual counterpart without risking the physical system~\cite{dietz2020integrating}. Eckhart and Ekelhart demonstrated a proof of concept, using a digital twin to detect a man-in-the-middle attack inside the virtual environment, to test the detection of violated security and safety rules~\cite{eckhart2018towards}.

\vspace{5px}
\subsubsection{Intrusion Detection}
DTs are increasingly being used to enhance the security of CPS by tapping into their anomaly detection ability to detect intrusions in systems. Here, we review studies that utilize DTs to address intrusion detection and cyberattacks in CPS. Akbarian~\el~proposed a digital twin-based intrusion detection system that leverages advanced techniques such as Kalman filter and support vector machine for attack detection and classification~\cite{akbarian2020intrusion}. Balta~\el~proposed a DT framework designed to detect cyberattacks in cyber-physical manufacturing systems by monitoring both controlled transient behaviors and anticipated process anomalies using run-time data~\cite{balta2023digital}. El-Hajj used DT with an intrusion detection system to secure physical devices in CPS, effectively detecting Hping3 attacks while also showing how different cyberattacks affect specific parts of a system~\cite{el2024leveraging}. By understanding these effects, designers are able to create security defenses that are better suited to the system's architecture and workload. Gehrmann and Gunnarsson introduced a DT-based security architecture to protect industrial automation and control systems (IACS) when opening up IACS low-level control functions and data exchange~\cite{gehrmann2019digital}. 

\vspace{5px}
\subsubsection{Policy Generation}
Traditionally, security policies are defined manually by domain experts to provide protection for an organization's communication and computing infrastructures. As infrastructure update cycles accelerate and cyberattacks grow in complexity, manually defined policies struggle to keep pace with emerging threats. To address this challenge, Hammar and Stadler proposed DTs as a solution for automating the development of effective security policies by simulating security scenarios in a controlled, virtual environment. Utilizing the data collected from simulations, reinforcement learning models are trained to automatically derive robust and adaptive security policies~\cite{hammar2023digital}. This approach not only enhances policy accuracy but also reduces human error and response time in rapidly evolving threat landscapes. 

The integration of digital twins into CPS offers a proactive and scalable solution to evolving security challenges due to their predictive, diagnostic, and simulation capabilities, allowing for effective security analysis, real-time intrusion detection, and policy generation without endangering the physical infrastructure. The reviewed studies demonstrate the versatility of DTs in detecting emerging threats and guiding adaptive defense strategies, making them increasingly essential in securing future CPS environments.

\subsection{Supply Chain}
The increase in counterfeit electronic parts poses significant reliability and security threats to the electronic supply chain supporting critical applications in defense, aircraft, vehicles, and medical devices due to their substandard quality, leading to major concerns for both government and industry~\cite{guin2014counterfeitProc,mark2015counterfeit}. As most electronic devices are produced with limited oversight and pass through insecure supply chains, they are vulnerable to malicious compromise. For example, an adversary can tamper with the device and create a backdoor, which can be exploited for malicious purposes. When deployed in systems, the compromised devices can jeopardize the entire infrastructure. Note that counterfeit electronics are non-authentic electronics originating from the recycling of used components from discarded electronics, defective/out-of-spec parts, overproduced, cloned, forged documentation, tampered, and remarking of low-grade components to their higher-grade counterpart (e.g., commercial-grade components as military-grade components). 


\vspace{5px}
\subsubsection{Root Cause Analysis}
One of the key advantages of DTs is their diagnostic capability, which allows for root cause analysis based on data collected throughout the system's lifecycle. This idea can be extended to the electronics supply chain to investigate the cause of electronics failures.
In~\cite{shaikh2022digital}, Shaikh \el mentioned that data collected during traditional design, verification, validation, and testing phases can be used to trace the root cause of faults, such as accelerated failure of new chips. The authors proposed a DT framework that is capable of detecting defective chips based on the data gathered throughout the electronics' lifecycle, as the DT is capable of looking at data associated with defective/out-of-spec chips and related data such as master results and wafer results records in the standard test data format 
database to determine if there is an observed anomaly. This framework uses causality analysis to provide traceability, thereby enabling comprehensive, end-to-end security management across the lifecycle.

\vspace{5px}
\subsubsection{Counterfeit Detection}
Hornik and Rachamim explored the application of DT and blockchain in what they referred to as Counterfeit Digital Twins (CoDT) to detect and prevent counterfeit electronics in/from the supply chain~\cite{Hornik2024codt}. The authors mentioned the limitations of traditional counterfeit detection methods, which often operate in isolation and lack real-time data sharing. As a solution, they proposed CoDT, as a primary data fusion technology to aggregate multiple counterfeit detection measures and technologies to more effectively detect counterfeits in real-time and provide recommendations to stakeholders, with the CoDT providing a secure and transparent means of tracking and monitoring the entire history of the electronics and verifying their authenticity using technologies such as CoDT-enabled blockchain, RFID and machine learning. The authors mentioned that the framework can verify the authenticity of electronics by comparing recorded data with actual product attributes to detect discrepancies. 

With the rise of counterfeit electronics posing serious reliability and security risks to critical systems, DTs can offer a promising solution to enhance traceability and transparency across the supply chain. Using data collected throughout the electronics lifecycle, DTs can provide insight into the root cause of component failures and support the detection of anomalies and unauthorized modifications. These capabilities can position DTs as a powerful tool for safeguarding electronic systems and ensuring the authenticity and integrity of devices from production to deployment.

\subsection{IoT Devices}
IoT devices are generally not equipped to run strong encryption mechanisms to secure the data they transmit~\cite{el2024systematic} and also suffer from weak authentication mechanisms due to their limited computing resources, which make them susceptible to unauthorized access. Below is a review of some digital twins-inspired solutions under three headings, namely: data privacy and security, threat detection, and security management.

\vspace{5px}
\subsubsection{Data Privacy and Security}\label{sec:data-privacy}

Low-powered IoT devices are often constrained by limited computational resources, making them susceptible to security breaches and privacy violations. To address this, authors in~\cite{cathey2021edge} proposed a novel approach that employs multiple digital twins for a single physical device. By segmenting data across distinct virtual representations using assigned tags, this method enables fine-grained access control, ensuring that only authorized users or applications can access specific data points, thereby preventing unauthorized data access.

\vspace{5px}
\subsubsection{Threat Detection}
An approach to enhance the performance of intrusion detection systems in IoT networks was proposed by Alharbi~\el~leveraging AI, DTs, and Blockchain technologies. In this framework, AI is utilized for real-time anomaly detection, DT replicates device behavior to predict potential threats, and Blockchain ensures secure and decentralized data transmission, collectively strengthening the overall security of the IoT networks~\cite{alharbi2024data}. Pirbhulal~\el~proposed a conceptual DT framework to enhance cybersecurity in IoT-based healthcare systems by creating a virtual replica of the targeted systems, enabling the identification of security vulnerabilities and potential breaches~\cite{pirbhulal2022towards}. To combat the threats facing Edge of Things (EoT) networks, Yigit introduced a digital twin-empowered smart attack detection system for 6G EoT networks, which integrates DT technology and edge computing to enable real-time monitoring and proactive threat detection, bolstering the security of IoT environments~\cite{yigit2023digital}. The system uses an online learning module to ensure continuous improvement by updating feature selection and classification methods, making it adaptable to dynamic attack landscapes.

\vspace{5px}
\subsubsection{Security Management}
DTs are a promising solution for improving security management in IoT environments. Empl~\el~proposed the SOAR4IoT framework, which integrates DTs with security orchestration, automation, and response (SOAR) to improve IoT security management through real-time monitoring, automated playbook-based responses, and enhanced coordination of security tools, effectively addressing the scale and complexity of IoT environments while reducing manual effort and human error~\cite{empl2022soar4iot}. In another work, Empl~\el~proposed the creation of a digital twin of the IoT network to enable proactive security management. This approach allows security analysts to continuously monitor the virtual representation of the network, detect potential threats early, and assess the impact of new configurations and updates to the network without putting the physical network at risk~\cite{empl2023digital}. Pittaras and Polyzos introduced a framework that uses DTs as a secure intermediary layer, enabling sensing and actuation through distributed ledger technologies~\cite{pittaras2022secure}. The authors mentioned that the consumers can interact with the physical devices solely through their digital representations, which were implemented with a smart contract, thereby minimizing direct exposure of physical devices and reducing attack surfaces. 

The application of DTs in securing IoT devices presents transformative opportunities in addressing critical challenges of low-powered devices, such as data privacy, intrusion detection, and security management. Therefore, leveraging DTs to safeguard IoT environments deserves greater attention and further exploration by the research community. 

\subsection{Cryptographic Systems}
Besides the mathematical security of cryptographic algorithms, the security of cryptographic systems is also affected by their implementations. A poorly implemented cryptographic system introduces side-channel leakage vulnerabilities, allowing sensitive assets to be leaked during encryption or decryption processes. Digital twins represent a promising solution to mitigate these risks through the simulation, monitoring, and analysis of side-channel behavior of the system in a virtual environment without interfering with the physical system. In the following, we describe two DT-based approaches that address information leakage in cryptographic systems based on recent studies.

\vspace{5px}
\subsubsection{Side Channel Analysis 
} 
Side-channel attacks exploit the implementation flaws in cryptographic algorithms to expose secret keys by observing power consumption, electromagnetic emissions, and execution time during cryptographic operations. To address this issue, Yi~\el~proposed a DT framework for side-channel resistant cryptographic devices by simulating the side channel information generated during the encryption process of cryptographic devices. By testing and analyzing the simulation model, the relevant side channel information and potential leakage points are captured and addressed. The authors tested different encrypted plaintexts, using a proposed measurement algorithm to accurately determine whether there is any side channel leakage~\cite{yi2023experimental}.

\vspace{5px}
\subsubsection{Thermal Analysis}
Thermal attacks are another subtle form of side-channel attack in which attackers may infer cryptographic keys or other sensitive information by analyzing the heating and cooling patterns of the system. To counter this, a thermal digital twin (TDT) solution was proposed to defend against thermal attacks with a focus on 3D System-in-Package (3D-SiP)~\cite{benelhaouare2024enhancing}. The authors utilized a high-fidelity simulation model that combines real thermal data with finite element modeling to analyze and monitor the thermal behavior of 3D-SiP systems in real time to enable proactive thermal risk management and improve system reliability and security. The application of DTs to develop side-channel resistant cryptographic systems is a promising approach to mitigating information leakage and thus deserves further investigation.

\subsection{Comparative Assessment} \label{subsec:comp-assessment}

\begin{table*}
\begin{threeparttable}
\centering
\caption{A Summary Study of Digital Twins for Security.}
\label{tab:security}
\small
\begin{tabular}{|P{1.4cm}|P{1.6cm}|P{0.7cm}|P{3.2cm}|P{1.2cm}|P{2.4cm}|P{0.6cm}|P{3.4cm}|}
\hline
\textbf{Domains} & \textbf{Scope} & \textbf{Refs.} & \textbf{Datasets [Availability]} & \textbf{Accuracy} & \textbf{Technologies} & \textbf{TRL} & \textbf{Limitations} \\
\hline

\multirow{12}{*}{CPS} 
& \multirow{7}{*}{Security} \multirow{7}{*}{Analysis} & \cite{shitole2021real} & Simulation data [N/P] & N/A & Cloud computing, IoT & M & Real-world validation \\
\cline{3-8}
& & \cite{dietz2020integrating} & Simulated network traffic between PLCs [P] & N/A & Sensors, Networking & M & Lacks performance evaluation, Cyber Threat 
integration \\
\cline{3-8}
& & \cite{eckhart2018towards} & TCP/IP network traffic [N/P] & N/A & Networking & M & Real-world validation, pipeline automation \\
\cline{2-8}
& \multirow{6}{*}{Intrusion} \multirow{6}{*}{Detection} & \cite{balta2023digital} & Run-time Data (temperature, motor counts) [P] & N/A & AI/ML, IoT & M & Detecting concurrent attacks and anomalies \\
\cline{3-8}
& & \cite{el2024leveraging} & Simulation data (CPU, Memory, Network Usage)[A/R] & 100\% & Data analytics & M & Real-world validation, Automated defense mechanism \\
\cline{3-8}
& & \cite{akbarian2020intrusion} & Simulation data [N/P] & $>83$\% & Cloud computing & M & Real-world validation \\
\cline{2-8}
& \multirow{3}{*}{Policy Gen.} & \cite{hammar2023digital} & Network traffic [N/P] & N/A & Networking, AI/ML & M & Generalizability to heterogeneous IT infrastructures \\
\hline

\multirow{2}{*}{\parbox{1cm}{Supply Chain}}
& \multirow{4}{*}{Counterfeit} \multirow{4}{*}{Detection} & \cite{shaikh2022digital} & N/A & N/A & AI/ML & L & Data accessibility \\
\cline{3-8}
& & \cite{Hornik2024codt} & N/A & N/A & RFID, Blockchain, AI/ML & L & Empirical testing and evaluation \\
\hline

\multirow{8}{*}{\parbox{1cm}{~~~IoT \\Devices}}
& Data Privacy  & \cite{cathey2021edge} & Simulation data [N/P] & 99\% & Networking, IoT & M & Interoperability across heterogeneous IoT platforms \\
\cline{2-8}
& \multirow{5}{*}{Threat} \multirow{5}{*}{Detection} & \cite{alharbi2024data} & N/A & N/A & AI/ML, Blockchain, IoT & L & Lacks performance evaluation, hardware implementation \\
\cline{3-8}
& & \cite{pirbhulal2022towards} & N/A & N/A & AI/ML, Big data, IoT & L & Real-world validation \\
\cline{3-8}
& & \cite{yigit2023digital} & Edge-IIoTset, ToN-IoT [P] & $>79$\% & AI/ML, Cloud computing, IoT & M & High computational overhead \\
\cline{2-8}
& \multirow{7}{*}{Security} \multirow{7}{*}{Manage-} \multirow{7}{*}{ment} & \cite{empl2022soar4iot} & Sensor data, Cyber threat intelligence data (NVD, CVE) [P] & N/A & Cloud computing, 
IoT & M & Limited scalability assessment \\
\cline{3-8}
& & \cite{empl2023digital} & Simulated network data [N/P] & N/A & IoT, Data analytics, Networking & M & Scalability, data availability \\
\cline{3-8}
& & \cite{pittaras2022secure} & Configuration and interaction data [N/P] & N/A & IoT, Blockchain  & M & Real-world validation \\
\hline

\multirow{2}{*}{\parbox{1cm}{Crypto-system}}
& SCA & \cite{yi2023experimental} & Simulation data (power traces) [N/P], DPA contest v4 [P] & N/A & AI/ML & M & Generalization and scalability \\
\cline{2-8}
& Thermal Analy. & \cite{benelhaouare2024enhancing} & Real thermal data [N/P] & N/A & Semiconductor packaging & M & High computation \\
\hline

\end{tabular}
\begin{tablenotes}
\item \hfill P: Published; A/R: Available on request; N/P: Not published; N/A: Not available; L: Low; M: Medium.
\end{tablenotes}
\end{threeparttable}
\end{table*}

This section provides a comprehensive comparative assessment of existing DT-based security studies across various domains. Specifically, we analyze each study in terms of its performance metrics, enabling technologies, Technology Readiness Level (TRL), and dataset. Furthermore, we highlight limitations and open research challenges in this evolving field. The studies reviewed span several critical domains, each of which presents unique requirements and security concerns that influence the design and application of DT solutions.

Table~\mbox{\ref{tab:security}} provides a detailed summary of recent research efforts that utilize digital twin technologies to address security challenges across a range of application domains. Column 1 specifies the domain in which each study is conducted (e.g., CPS, supply chain, IoT devices, and cryptosystem), while Column 2 outlines the specific scope or focus of the study within that domain (see Figure~\mbox{\ref{fig:HWSec_Tax}}). Column 3 lists the corresponding references, pointing to the original research articles. Column 4 describes the types of datasets employed in these studies, which may include simulation-generated data, real-time operational data, or publicly available datasets such as those from the National Vulnerability Database (NVD) and Common Vulnerabilities and Exposures (CVEs). Column 5 presents the reported performance metrics, typically accuracy, where available (e.g., in~\mbox{\cite{akbarian2020intrusion}, \cite{el2024leveraging}}, etc.), and is marked as ``N/A'' when such data is not provided. Column 6 outlines the enabling technologies employed, including machine learning, cloud computing, IoT, and blockchain frameworks. Column 7 indicates the technology readiness level of each study, reflecting its degree of maturity: for simplicity, a low TRL corresponds to the conceptual research phase, medium indicates experimental validation, and high denotes deployment in real-world environments. Finally, Column 8 summarizes the major open challenges and limitations identified by each study, highlighting areas for future research and development in the field of DT-based security. 
While many of the reviewed studies demonstrate the promising potential of DT technology in enhancing security across domains such as IoT, CPS, and cryptographic systems, the majority remain at the experimental stage~\cite{shitole2021real,dietz2020integrating,eckhart2018towards,balta2023digital,el2024leveraging,hammar2023digital,akbarian2020intrusion,cathey2021edge,yigit2023digital,empl2022soar4iot,empl2023digital,pittaras2022secure,yi2023experimental,benelhaouare2024enhancing}. A smaller subset of works, including~\cite{shaikh2022digital,alharbi2024data,pirbhulal2022towards, Hornik2024codt}, remain in the conceptual phase with no reported real-world implementation or validation. Although applications in CPS and IoT have received comparatively more attention, including some hardware-based experimentation, these efforts are still constrained by challenges related to scalability, integration into existing infrastructures, and deployment in operational settings. Overcoming these limitations is essential for advancing DT-based security solutions from research prototypes to robust, industry-grade implementations.

\begin{figure*}[ht]
    \centering
    \includegraphics[width=1\textwidth, clip]{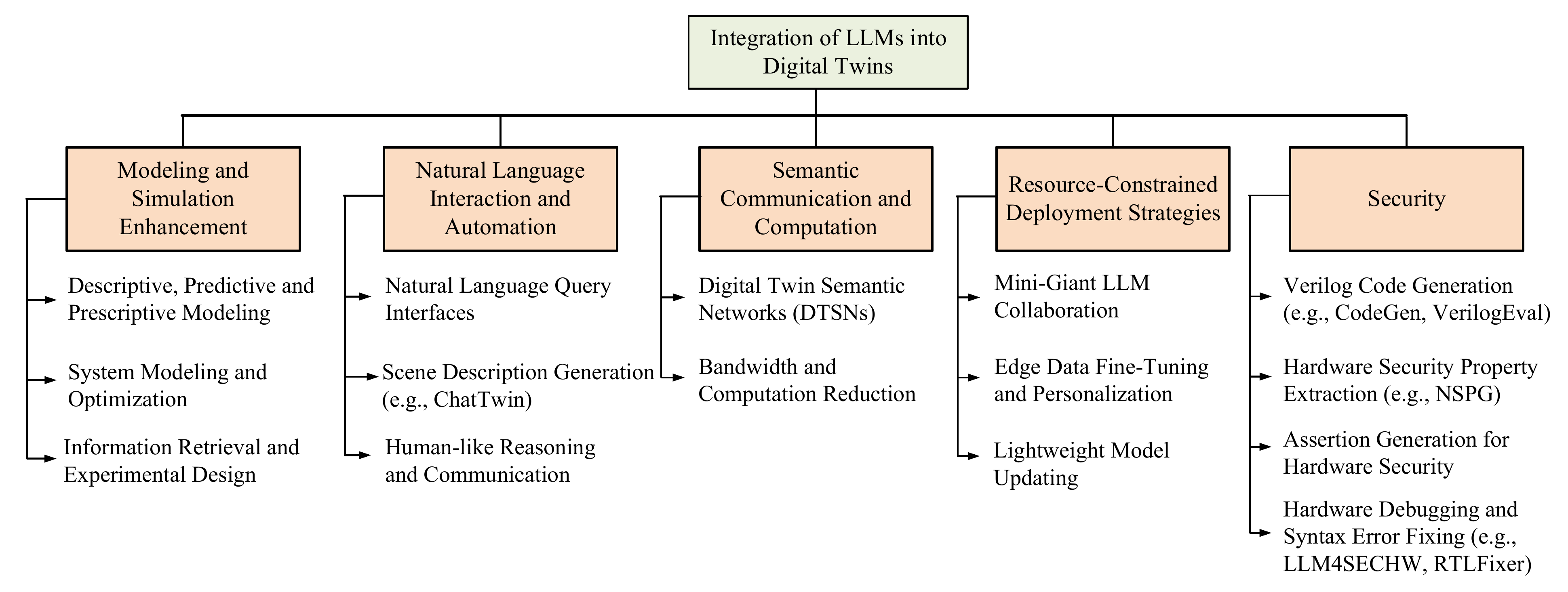} 
    \caption{\centering Taxonomy of the integration of LLMs in DT frameworks.
    }
    \label{fig:LLM_Tax}
\end{figure*}

\section{Large Language Models in Digital Twin Frameworks} \label{sec:llms}

The preceding sections have explored how DTs revolutionize physical system modeling by creating dynamic virtual replicas that mirror real-world environments. Large language models (LLMs), with their natural language processing capabilities, have demonstrated their potential to enhance the cognitive abilities of DTs. This convergence creates systems that not only replicate physical counterparts but also understand, communicate, and reason about them in human-like ways. The following discussion explores the evolution of LLMs, their broad applications across domains, and their specific implementations within DT frameworks. The following discussion contains an overview of the integration of LLMs in DT frameworks and, especially, security. With an objective to analyze the existing studies and their limitations, we outline a thorough survey methodology and anchor the use cases of LLMs in distinct DT flows.

\subsection{An Introduction to Large Language Models}\label{sec:llms-intro}
LLMs represent a groundbreaking advancement in AI, functioning as sophisticated deep learning systems capable of processing and generating human-like text. The evolution of these models began with the introduction of the `transformer' architecture in 2017, which marked a significant shift from traditional recurrent neural networks~\mbox{\cite{vaswani2017attention}}. Another significant milestone in LLM development is the introduction of bidirectional encoder representations from transformers (BERT) in 2018, which demonstrated unprecedented versatility through extensive training on natural language data~\mbox{\cite{devlin2019bert}}. OpenAI's GPT series marked another crucial evolution phase, with GPT-3's 2020 release followed by GPT-4 more recently, has further enhanced LLMs through instruction tuning and user-friendly conversational interfaces~\mbox{\cite{achiam2023gpt}}. LLMs have transformed numerous industries through their natural language processing capabilities, performing multiple language tasks - answering questions, summarizing documents, translating languages, and completing text sequences. This flexibility enables them to enhance traditional workflows across domains. Their remarkable ability to make predictions based on relatively small input prompts has enabled generative AI applications that produce human-like content in response to natural language instructions~\cite{naveed2025comprehensive}.

\subsection{Integration of LLMs in DT Frameworks}
From manufacturing to healthcare and urban planning, DTs generate vast amounts of data that can be difficult to interpret and utilize effectively. Integrating LLMs into these frameworks allows more intuitive interaction with DTs and augments explainable decision-making across multiple domains. Figure~\ref{fig:LLM_Tax} showcases the exploration of LLMs in DT systems so far, with the potential to go even beyond the following scopes:

\vspace{5px}
\subsubsection{Modeling and Simulation Enhancement}
LLMs can conduct descriptive, predictive, and prescriptive modeling by interpreting vast and complex datasets, thus holistically improving the accuracy and reliability of predictions. TWIN-GPT, for example, is fine-tuned on a pre-trained LLM on clinical trial datasets to generate personalized digital twins for different patients~\cite{wang2024twin}. LLMs also assist in system modeling and optimization through intelligent analysis of parameter interactions and system dynamics, and LLMs support information retrieval and experimental design by automatically synthesizing insights from extensive simulations, streamlining hypothesis testing and data-driven decision-making in complex simulation scenarios. For instance, a recent study proposes using LLM agents within a shopping mall twin to simulate human behaviors and thermal preferences~\cite{yang2024llm}. These LLM agents interact with the DT environment for optimizing HVAC systems.

\vspace{5px}
\subsubsection{Natural Language Interaction and Automation}
LLMs bring natural language understanding to DT systems and improve human-system interaction and automation. Natural language query interfaces allow users to communicate with digital twins without requiring deep technical expertise. LLMs can automate the generation of scene description documents for DTs, reducing the need for manual processes and domain expertise. ChatTwin, a conversational system that automates the generation of scene description documents for data center DTs, showcases the ability of LLMs to effectively model and generate data center scenes of the DT~\cite{li2023chattwin}. LLMs can also provide an explainability platform, generating natural language explanations of the system’s decision-making by utilizing domain-specific knowledge bases. Retrieval-augmented generation (RAG) allows LLMs to utilize technical documents, scientific papers, and code to explain real-time decisions taken by the system. 

\vspace{5px}
\subsubsection{Semantic Communication and Computation}
Semantic communication within digital twin frameworks is improved by integrating LLMs, notably through the concept of digital twin semantic networks by more efficient communication and computation within DT environments~\cite{hong2024llm}. This approach reshapes both intra-twin and inter-twin communication architectures through semantic-level frameworks that unify efficient communication and computation. By enabling DTs to communicate and compute at the semantic level rather than through raw data exchange, this framework significantly reduces bandwidth requirements and computational overhead.

\vspace{5px}
\subsubsection{Resource-Constrained Deployment Strategies}
Resource-constrained environments pose unique challenges for deployment, which can be effectively addressed through integration with LLMs. A mini-giant LLM collaboration scheme designed for resource-constrained environments involves edge data fine-tuning and instruction prompts, where larger models provide rich knowledge information while smaller models handle personalized expertise updating. Moreover, lightweight model updating techniques by LLMs enable continuous and efficient model refinement, ensuring up-to-date system intelligence without extensive computational resources.

\vspace{5px}
\subsubsection{Security}
Security is critically enhanced by incorporating LLMs into DT environments. LLMs can automatically generate secure hardware description code, improving accuracy and security at the hardware design level. Thakur~\el demonstrated how LLMs, such as CodeGen, can automate synthesizable Verilog code generation~\cite{thakur2024verigen} at the register transfer level (RTL). Advanced property extraction mechanisms, such as neural security property generation, utilize the semantic understanding capabilities of LLMs to reliably identify and extract hardware security properties. Furthermore, LLM-based assertion generation tools support proactive hardware vulnerability detection, while debugging and syntax error correction solutions enable faster resolution of security-critical errors. Kande~\el investigated LLMs for automatic generation of hardware security assertions~\cite{kande2023llm}. Their research explored the use of natural language prompts in LLMs to produce SystemVerilog assertions, which are indeed challenging to write manually. Tsai~\el introduced RTLFixer, a framework integrating LLMs to automatically identify and rectify syntax errors in RTL code~\cite{tsai2024rtlfixer}. This framework also notably utilizes RAG and an advanced prompting technique that results in exceptional proficiency in resolving syntax errors in code.

\subsection{Future Prospects of LLMs and DTs in Security}

Despite such promising advancements, integrating LLMs with DTs has introduced several critical issues along the way. Notably, in resource-constrained deployment scenarios where compute is limited, computational efficiency remains an area of concern. Another challenge is the accuracy and reliability of the LLM-generated outputs that require robust validation frameworks, especially within safety-critical domains like healthcare and manufacturing.

As these challenges are gradually acknowledged and dealt with, LLMs and DTs together truly represent a strong unified solution for advancing security measures across cyber-physical systems. With the logical reasoning and natural language understanding capabilities of LLMs, the DT frameworks in the future can offer improved security analysis and anomaly detection. Recent trends include training specialized models for more accurate security-related applications and exploring the multimodality of LLMs to interpret both textual and visual data to provide real-time security evaluations of DT environments. We anticipate that this convergence between LLMs and DTs will continue to significantly strengthen the security of critical infrastructure.

\section{Challenges and Limitations}\label{sec:DT-challenges} 

Despite recent studies showcasing the potential of DTs across various domains, as discussed in Section~\ref{sec:application}, due to their ability to create virtual representations of physical systems to carry out optimization, fault detection, enhanced security, and all forms of analysis without interfering with the physical systems, this promising technology is not without its challenges and limitations. In this section, we examine some of these challenges and limitations of this technology and present some of the solutions proposed in the literature.

\subsection{Implementation Cost}
The implementation of DTs demands significant technical expertise, resources, and investment. For the accurate representation of a physical system in the digital space, the DT must incorporate detailed modeling of the system’s structure, behavior, and interactions, which are often at multiple levels of abstraction. This includes the integration of high-fidelity physics-based models, real-time data from sensors, and AI-driven simulations that reflect real-time changes in the physical environment. Achieving this level of precision often requires specialized tools, high-performance computing infrastructure, and extensive domain knowledge, which can be difficult to assemble and maintain, making the DT more costly than the physical asset at times.

\subsection{Data Acquisition}
DTs rely heavily on high-quality, real-time data streams to accurately mirror the behavior and condition of their physical counterparts. However, missing data points, sensor noise, or inconsistencies in measurements can significantly degrade the accuracy and reliability of DT's predictions and decision-making capabilities. As a result, substantial effort must be invested into data cleaning, validation, and preprocessing to ensure that the DT operates as intended. This challenge is further compounded in the semiconductor industry due to data accessibility issues, as key stakeholders such as foundries, testing facilities, and equipment vendors may be reluctant to share proprietary or sensitive operational data due to intellectual property concerns, data privacy, or competitive reasons. This reluctance can hinder the construction of comprehensive DTs by creating blind spots in critical process stages or component behavior. Without full visibility into the system, the DT may be forced to rely on approximations or incomplete models, reducing its value in optimization and root-cause analysis. 

\subsection{Security and Privacy}
The continuous flow of data between the physical twin and its virtual counterpart introduces significant security and privacy risks. This data exchange creates entry points for cyberattacks, including denial of service, false data injection, and unauthorized access~\cite{saad2020implementation}. Attackers may exploit these vulnerabilities to manipulate data, falsify system states, or mislead predictive algorithms, and ultimately compromise the integrity and functionality of the digital twin.

The increased use of IoT devices further amplifies the security issues due to their limited built-in security features. IoT devices are resource-constrained devices, making them susceptible to firmware vulnerabilities, weak encryption, and weak authentication mechanisms. Once compromised, these devices can serve as gateways for attackers to infiltrate the broader digital twin infrastructure.

From a privacy perspective, the data collected and transmitted in healthcare, smart cities, or industrial operations can contain sensitive or proprietary information. Without the use of strong encryption and access control mechanisms, this information may be exposed to unauthorized entities, leading to undesired privacy breaches. Possible ways to address these challenges are the integration of blockchain as discussed in~\cite{huang2020blockchain,putz2021ethertwin,son2022design}. However, this may add an extra layer of complexity and latency to the system. 

\subsection{Token Size and Hallucination Problems of LLMs}

While LLMs are considered to have the relatively untapped potential to enhance the security of DTs, they are inherently constrained by a fixed token window that limits the amount of input data they can process in a single inference. This poses a significant challenge while analyzing DTs of complex hardware systems because large-scale system logs, event traces, or multi-modal telemetry data often exceed this limit. In such situations, the inputs get truncated and critical context gets omitted in the process. In security-sensitive applications, this might lead to generating incomplete or even misleading outputs.


Moreover, as probabilistic token predictors, LLMs are prone to generating information that is syntactically plausible but semantically incorrect. This is termed as hallucination. For example, when tasked with identifying side-channel vulnerabilities in a hardware digital twin, an LLM may hallucinate and fabricate leakage channels that do not even exist in the actual system. While this feature comes inherently with the non-deterministic nature of LLMs, it significantly reduces the reliability of the generated responses, especially in security.



\begin{figure}[t]
    \centerline{\includegraphics[width=1.05\linewidth]{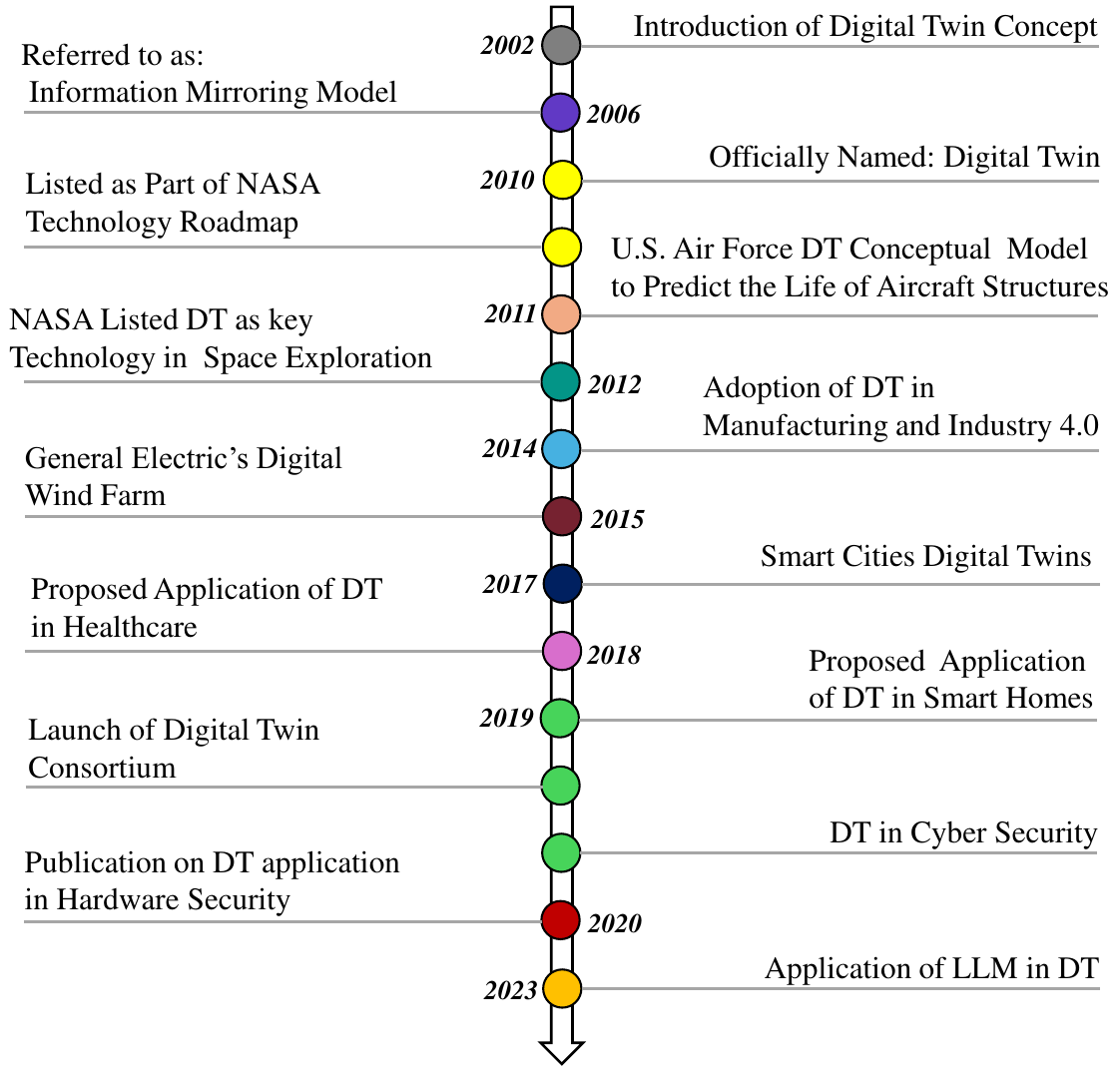}} 
    \caption{\centering Major milestones in the development and adoption of digital twin technology. } \vspace{-10px}
    \label{fig:DT-trend}
\end{figure}

\section{Research Trends} \label{sec:trens}
In the last two decades,  digital twin-related publications in conference proceedings, journals, or magazines have increased tremendously. The rapid growth in the number of publications in this field is due to the promising future of DTs and the advantages they offer in any given application. More recently, their use in enhancing the security of systems has emerged as a topic of growing interest within the research community, as discussed in Section~\ref{sec:DT-in-HS}, and this is expected to continue in the years to come.

Figure~\ref{fig:DT-trend} shows the major milestones in the development and adoption of digital twin technology. Although DT was introduced by Dr. Michael Grieves in 2002 as a framework for product lifecycle management, there were only a few research works on the topic until 2014, when it was adopted in the manufacturing industry. This was a turning point that marked the beginning of digital twins' practical relevance. Its widespread adoption in manufacturing and Industry 4.0 was catalyzed by the IoT and IIoT technologies that provided the necessary real-time data streams and sensor infrastructure to operationalize digital twin systems. This adoption in Industry 4.0 and the rise in IoT caused increased attention from researchers in this field. In 2015, `Digital Wind Farm' was announced by General Electric, which was one of the earliest commercial implementations of digital twin systems. Since then, several studies have been done by academia in industry to reduce downtime, optimize production lines, and perform predictive maintenance~\cite{tao2018digital, booyse2020deep, aivaliotis2019use}. In the last decade, research in the development and implementation of digital twins of cities, smart homes, and healthcare has received huge attention from the research community, leveraging the power of IoT and wearable devices. These developments highlight the versatility of digital twins in modeling dynamic, human-centric, and infrastructure-rich environments.
Since 2019, we have experienced an increased number of publications on the use of DT to solve some cybersecurity problems, such as intrusion detection. This number is expected to keep rising in the coming years. 
Since 2020, researchers in hardware security have begun exploring the use of digital twin frameworks to address critical challenges such as information leakage, fault injection, and hardware counterfeiting~\cite{shaikh2022digital, dietz2020integrating, Hornik2024codt, empl2023digital, yi2023experimental}. While this emerging line of research holds significant promise, it remains in its early stages as most of the existing studies are conceptual in nature, with practical implementations being limited. 

These milestones are not merely historical footnotes; they reflect the maturation of digital twins from an abstract concept into a multidimensional technology with real-world impact. Each turning point — whether the entry into manufacturing, healthcare, or security — has incrementally extended the capabilities and scope of DT systems. As the fusion of digital twins with LLMs becomes more prominent, we are likely to witness the emergence of intelligent, adaptive DTs capable of semantic reasoning, self-optimization, and autonomous decision-making. The convergence of these technologies represents a pivotal shift, potentially transforming DTs from passive monitoring systems into active, explainable, and secure agents across critical domains.

\section{Conclusion} \label{sec:conclusion}
Digital twins are rapidly gaining attention, and their applications in security demonstrate enormous potential. Therefore, the integration of DTs into security frameworks is critical and demands focused attention. In this paper, we reviewed how DTs improve performance, reliability, and security across key domains, such as energy, smart homes, transportation, healthcare, and manufacturing. We presented, for the first time, a unified study that surveys recent digital twin–based security applications across cyber-physical systems, the Internet of Things, and cryptographic systems, focusing on security use cases of digital twins for counterfeit electronics detection, intrusion detection, and information leakage prevention. We also summarized the integration of LLM with DT for security workflows, the challenges and limitations of DT applications in hardware security, and highlighted the ongoing research trends, concluding with a future roadmap for advancing DT-based hardware security solutions. 

\section*{Acknowledgment} This work was supported by the National Science Foundation under Grant Number CNS-2312139.

\bibliographystyle{ieeetr}
\bibliography{guin-bib,other-bib}
\begin{IEEEbiography}[{\includegraphics[width=1in, clip, keepaspectratio]{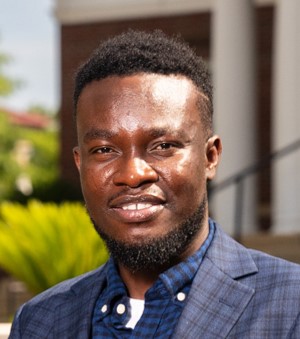}}]{Blessing Airehenbuwa (S'22)} received his M.S. in Electrical and Computer Engineering from Tuskegee University, Tuskegee, AL, USA, in 2024. He is currently pursuing his Ph.D. in Electrical Engineering from the Department of Electrical and Computer Engineering, Auburn University, AL, USA. He earned his B. Eng. in Electrical/Electronics Engineering from the University of Benin, Benin City, Nigeria, in 2016. His research interests include hardware security, VLSI design \& test, cybersecurity, and digital twins. He is a student member of the IEEE.
\end{IEEEbiography}
\vspace{-30px}
\begin{IEEEbiography}[{\includegraphics[width=1in, clip, keepaspectratio]{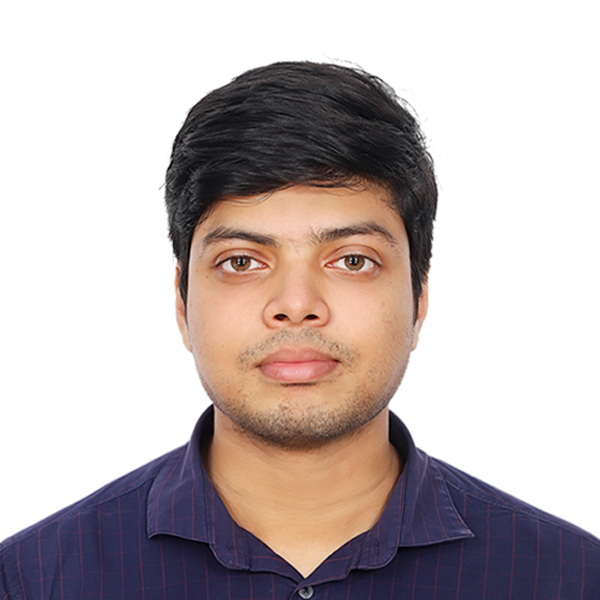}}]{Touseef Hasan (S'25)} is currently pursuing his Ph.D. from the Department of Electrical and Computer Engineering in the School of Computing at Wichita State University, Wichita, KS, USA. He received his B.Sc. in Naval Architecture and Marine Engineering from Bangladesh University of Engineering and Technology, Dhaka, Bangladesh in 2024. His research interests include exploring the practicality and accessibility of LLMs. He is a student member of the IEEE.
\end{IEEEbiography}
\vspace{-30px}
\begin{IEEEbiography}[{\includegraphics[width=1in, clip, keepaspectratio]{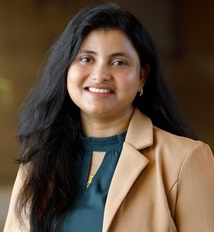}}]{Souvika Sarkar (M'25)} received her Ph.D. in Computer Science and Software Engineering from Auburn University in 2020. She is currently an Assistant Professor in the School of Computing at Wichita State University, Wichita, KS, USA. Prior to that, she completed her B.Tech in Computer Science and Engineering from West Bengal University of Technology, India, in 2012, and earned her M.E. in Software Engineering from Jadavpur University in 2018.

Dr. Sarkar’s research interests include natural language processing (NLP), generative AI, LLMs, and building scalable and adaptive AI systems. Her research centers on developing efficient NLP systems, with a strong focus on interpretability, adaptability, and real-world impact. She has published widely in top-tier venues in AI and NLP. She actively serves the research community as a reviewer for leading conferences, including ACL, EMNLP, NAACL, and AACL.
\end{IEEEbiography}
\vspace{-30px}
\begin{IEEEbiography}[{\includegraphics[width=1in, clip, keepaspectratio]{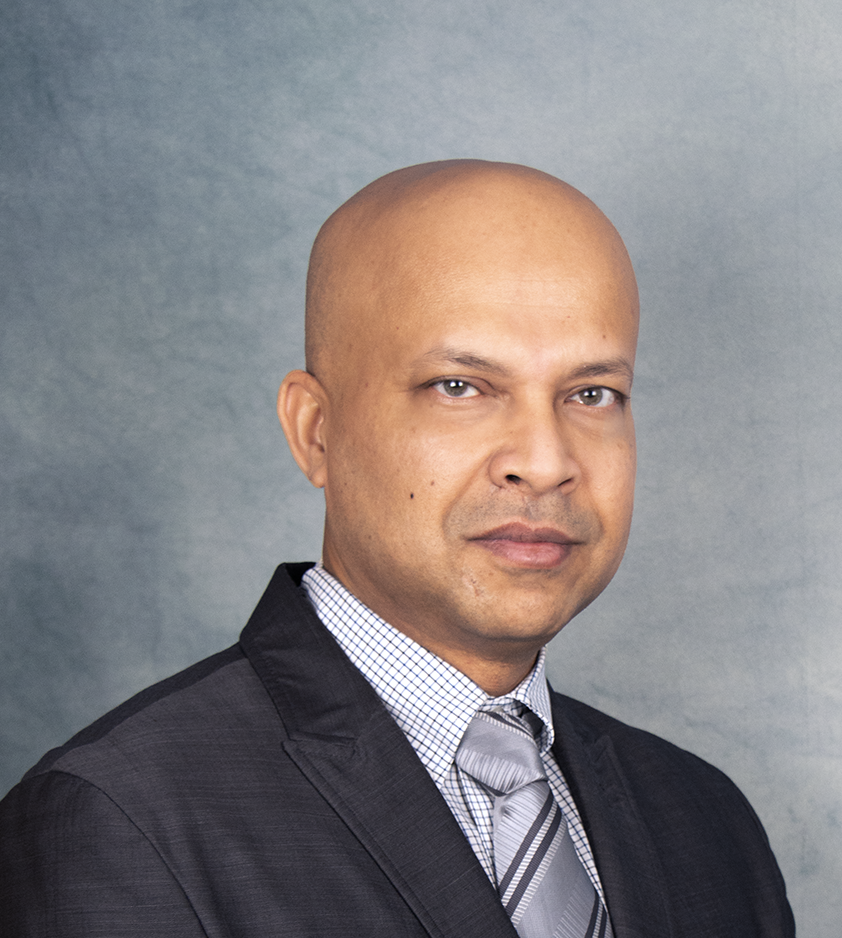}}]{Ujjwal Guin (S'10--M'16--SM'22)} received his Ph.D. in Electrical and Computer Engineering from the University of Connecticut in 2016. He is currently the Godbold Associate Professor in the Department of Electrical and Computer Engineering at Auburn University, Auburn, AL, USA. He earned his B.E. in Electronics and Telecommunication Engineering from Bengal Engineering and Science University, India, in 2004, and his M.S. in Electrical and Computer Engineering from Temple University, Philadelphia, PA, USA, in 2010.

Dr. Guin's research interests lie in the areas of hardware security and trust, supply chain security, cybersecurity, and VLSI design and test. He has developed a range of on-chip structures and techniques aimed at enhancing the security, trustworthiness, and reliability of integrated circuits. He has authored thirty-plus journal articles and over fifty conference papers, with several earning best paper nominations and awards. His research has been supported by the National Science Foundation (NSF) and several DoD institutions. He actively contributes to the academic community by serving on organizing and technical program committees of leading conferences, including HOST, ITC, VTS, PAINE, and several other well-regarded venues.
\end{IEEEbiography}

\balance

\end{document}